# Non-equilibrium and non-linear stationary state in thermoelectric materials

H. Iwasaki[a], M. Koyano[a], Y. Yamamura[b] and H. Hori[a]

[a]School of Materials Science, JAIST, Tatsunokuchi 923-1292, Ishikawa, Japan

[b]Center for Nano Materials and Technology, JAIST, Tatsunokuchi 923-1292, Ishikawa, Japan

Efficiency of thermoelectric materials is characterized by the figure of merit $Z$. $Z$ has been believed to be a peculiar material constant. However, the accurate measurements in the present work reveal that $Z$ has large size dependence and a non-linear temperature distribution appears as stationary state in the thermoelectric material. The observation of these phenomena is achieved by the Harman method. This method is the most appropriate way to investigate the thermoelectric properties because the dc and ac resistances are measured by the same electrode configuration. We describe the anomalous thermoelectric properties observed in mainly $(Bi,Sb)_2Te_3$ by the Harman method and then insist that $Z$ is not the peculiar material constant but must be defined as the physical quantity dependent of the size and the position in the material.

The figure of merit Z was proposed as a character which represents an efficiency of thermoelectric materials and was fundamentally determined in the condition of a finite temperature difference $T_h$-$T_c$ between the high temperature side and the low temperature one in the material. Assuming the temperature dependences of the resistivity $\rho$, the Seebeck coefficient $\alpha$ and the thermal conductivity $\kappa$ to be negligible on the temperature range of $T_h$-$T_c$ or to be replaced by the average of the values at the heating and the cooling ends of the material, Z is usually represented by $Z=\alpha^2/\rho\kappa$ (K$^{-1}$). Z has been conventionally believed to be a sample size independent material constant. This means that Z and the temperature gradient, which is induced by the dc current, are constant in a sample. Three physical properties of $\alpha$, $\rho$ and $\kappa$ are measured to estimate Z, while T. C. Harman proposed another method[1] in which the only resistance measurements by the ac and dc methods are required. The dimensionless figure of merit ZT is given by $ZT=(R_{dc}-R_{ac})/R_{ac}$ equivalent to the above formula of Z, where $V_{dc}=R_{dc}I=R_{ac}I+\alpha\Delta T$ and $R_{ac}$ by the ac method corresponds to the ohmic resistance of a sample. ZT is also rewritten by $ZT=(\alpha/\rho j)(\Delta T/L_V)$, where j is the current density and $\Delta T$ and $L_V$ are the induced temperature difference and the length between the voltage terminals, respectively. This expression indicates that the temperature difference in the sample is possible to be studied by the dc measurements using the voltage terminals as the thermal probe. The temperature distribution in the sample can be determined by the measurements in several dc voltage terminals lengths. Thus, the Harman method can be applied to not only the evaluation of Z but also the determination of the temperature distribution inside the sample, in other words, positional dependence of Z.

The fabrication for the evaluation of ZT of thermoelectric materials has already reported in detail[2]. The Harman method is also applicable to the evaluation of thermoelectric modules[3]. In the Harman method a sufficient adiabatic condition of a sample is quite important and its condition must be satisfied well in the experiment. The experiments are made on establishment of the satisfactory adiabatic condition of the sample. It should be emphasized that the obtained ZT values were exactly the same even if the $\Delta T$ induced by the dc current in the sample was changed by 100 times from

~10mK to ~1K[2]. The influence of a heat-leakage from a heat bath including the thermal radiation from the lateral face of the sample was negligible to the obtained results in our equipment. The independence of ZT in the induced $\Delta T$ is especially important point in the investigation of the temperature distribution inside the sample discussed below. In the present method the temperature difference in the sample is ~100-200mK and the measurements are made near the limit of $\Delta T$~0. Therefore, the expression of $Z=\alpha^2/\rho\kappa$ is quite reliable to our experiments contrary to the usual evaluation method. The same value of the dc resistivity (or ZT) actually observed in the different $\Delta T$ in the sample[2] also indicates that ZT can be constant in the temperature difference at least below 1K. Sintered samples were synthesized by a hot press method ($(Bi,Sb)_2Te_3$) and a spark-plasma sintering one ($Co_{0.9}(Pt,Pd)_{0.1}Sb_3$ and $\beta$-$Zn_4Sb_3$). Three p-type $(Bi,Sb)_2Te_3$ samples (#1-1, #1-2 and #2) are prepared. The samples with several lengths denoted by #1-1 are cut out of the same batch and the cross section is commonly 1.9mm$^2$[2]. The sample #1-2 is also prepared in the same way and has the length of 15mm and the cross section of 0.53mm$^2$. The cross section in a series of the samples #2 is 1.6mm$^2$. The #1 and #2 samples have the different Hall coefficient.

Figure 1 shows the sample length L dependences of ZT in the samples of #1-1 and #2 of $(Bi,Sb)_2Te_3$ at 300K. The length between the voltage terminals is common of $L_V$=2mm except for $L_V$=1mm in the samples with L below 4mm. In both samples ZT increases largely and linearly with decreasing L and its gradient is larger in the sample with small cross section than in the sample with large one. The ZT values in the samples with short length coincide well with those by the conventional method evaluated by $Z=\alpha^2/\rho\kappa$, where the thermal conductivity is also measured in the sample with short length. We interpret that ZT of the material depends on the sample size in itself because ZT changes largely by 30-40% with the sample length although the measurements are performed in the sufficient adiabatic condition as emphasized before and the non-linear stationary temperature distribution is observed in the sample described below.

Next we show the results of the voltage terminal length $L_V$ dependence on the sample #1-2 with the Seebeck coefficient of 205μV/K at 300K. Two voltage terminals are set at the positions of symmetry to the center of the sample for each $L_V$. The temperature dependences of the resistivity for various lengths $L_V$ are given in Fig. 2. The dc resistivity values decrease rapidly near the ends of the sample and the small decrease is observed in the short voltage terminal lengths below $L_V$=9.16mm. These phenomena are always reproducible in the other $(Bi,Sb)_2Te_3$ samples. The observed $L_V$ dependence in the dc resistivity indicates that the temperature difference between the voltage terminals decreases largely near the ends of the sample. The temperature distribution in the sample is confirmed to be stable because the dc resistivity does not change even if the wait time in the dc voltage measurements is changed from 4.5 minutes to 25 minutes, where wait time is necessary to measure a saturate voltage value in the steady heat-flow for each dc current direction. These time independent dc resistivities are observed for every $L_V$. Using the obtained dc and ac resistivities and the relation of $ZT=(\alpha/\rho j)(\Delta T/L_V)$, the temperature difference in the sample is estimated. The results are shown in the inset of Fig. 2 and $\Delta T$ versus $L_V$ obviously deviates from the linear relation above $L_V$=9mm. The obtained $\Delta T$ corresponds to the average value between the voltage terminals and the real temperature gradient at each position is expected to be much larger especially near the ends of the sample. It is concluded that this anomalous non-linear temperature distribution in the sample exists as a steady state and the phenomena induced by the dc current is essentially non-equilibrium in the thermoelectric material $(Bi,Sb)_2Te_3$. Possibility of the non-linear temperature distribution has been pointed out[4] which is caused by the thermal radiation due to the temperature difference between the heat bath and the heated and cooled ends of the sample induced by the dc current. However, this effect is discarded here because of the same value of ZT obtained in the different temperature difference in our equipment[2]. The negligible thermal radiation is also supported by the fact that ZT is much smaller in the sample #1-2 with small lateral face than that in the #1-1 sample as discussed later. The authors in ref. 4 seem to take notice of the size effect of Z because the length and the cross section of sample are quite different from those in ref. 1 and interpreted the effect to be due to the thermal radiation from the lateral face. The linear positional dependence of the temperature was only reported there.

In order to confirm the validity of the observed phenomena, the following two kinds of measurements are performed. The first is the estimation of the amplitude in the thermal heat-flow by the low frequency ac current which makes a cross check to the temperature distribution in the sample. The procedure in which the voltage terminals are used as the thermal probe is also applicable. The voltage between the voltage terminals by the low frequency ac current is given by summation of the ohmic voltage and the thermoelectric power. When the ac current with a sine-wave is applied, the voltage, $V(\omega t+\Delta\phi)=V_{ohm}(\omega t)+V_{thermal}(\omega t+\Delta\phi_0)=R_{ohm}I_0\sin(\omega t)+\alpha\Delta T_0\sin(\omega t+\Delta\phi_0)$ appears. The second term corresponds to the thermal heat-flow which phase is delayed by $\Delta\phi_0$ against the current. The ohmic voltage is determined experimentally and the thermal heat-flow is obtained by the fitting of the measured voltage as the adjustable parameters of amplitude $\alpha\Delta T_0$ and phase delay $\Delta\phi_0$. The observed voltages in the sample #1-2 are given in Fig. 3, where the frequency is $f=\omega/2\pi=5mHz$. The obvious phase delay $\Delta\phi$ is seen in the measured voltage for all length $L_V$. The amplitude of the thermal heat-flow is shown as a function of $L_V$ together with the amplitude of the ohmic voltage $R_{ohm}I_0$ in the inset. The obtained amplitude of the thermal heat-flow increases linearly with increase of the voltage terminal length and then deviates from the linear relation near the ends of the sample. The position, at which the amplitude of the thermal heat-flow begins to deviate, is approximately 9mm and is consistent with the previous results. The details including the phase delay and the frequency dependence will be discussed elsewhere.

The second is the comparison of ZT (or $\Delta T$) behavior among thermoelectric materials. The samples of $Co_{0.9}(Pt,Pd)_{0.1}Sb_3$ and $\beta$-$Zn_4Sb_3$ are used here and the sample length is 10mm. The $L_V$ dependences of the estimated ZT values are summarized in Fig. 4. In these samples including $(Bi,Sb)_2Te_3$ #1-2 the ZT values are quite small as discussed in Fig. 1 because the sample length is 10-15mm. We must also make a comment that ZT is affected by the cross section, that is, the value of ZT=0.435 in the $(Bi,Sb)_2Te_3$ #1-2 sample is much smaller than those in the samples #1-1 as given in Fig. 1. Therefore, ZT depends on not only the sample length but also the cross section of the sample. The anomaly of ZT near the ends of the samples is quite small in both materials in contrast to that of $(Bi,Sb)_2Te_3$. The degree of the anomaly of ZT, $\Delta(ZT)/(ZT)_{max}$ is 0.058 ($(Bi,Sb)_2Te_3$), 0.018 ($Co_{0.9}(Pt,Pd)_{0.1}Sb_3$) and 0.041 ($\beta$-$Zn_4Sb_3$), where $\Delta(ZT)$ is the difference between the minimum $(ZT)_{min}$ near the center of the sample and the maximum $(ZT)_{max}$ near the ends of the sample. The large value of $\Delta(ZT)/(ZT)_{max}$ is not originated to the experimental condition but is interpreted to be intrinsic in $(Bi,Sb)_2Te_3$. This is obvious by taking account of the facts that the ZT value is independent of the induced $\Delta T$ by changing applied dc current in the same sample[2] and that $\Delta(ZT)/(ZT)_{max}$ is quite different between the $Co_{0.9}(Pt,Pd)_{0.1}Sb_3$ and $\beta$-$Zn_4Sb_3$ materials though the ZT value is close to each other. Thus, we conclude that the temperature gradient in the sample induced by the dc current is not uniform and is larger near the ends of the sample. These phenomena are obvious in the $(Bi,Sb)_2Te_3$ material with relatively large ZT values. The large ZT values in the samples with short length in Fig.1 are basically attributed to the existence of the large temperature gradient only near the ends of the sample. In other words, the relatively small ZT value in the samples with large length is caused by the existence of small temperature gradient around the center in it. The non-equilibrium and non-linear stationary state appears in the condition of flowing dc current and is more obvious in the samples $(Bi,Sb)_2Te_3$ ($\kappa$~13mW/Kcm at 300K) and $\beta$-$Zn_4Sb_3$ ($\kappa$~11mW/Kcm at 300K) with small thermal conductivity. The phenomena may be related to thermal transport in the materials.

In the present study it is made clear that Z defined by $\alpha^2/\rho\kappa$ ($K^{-1}$) depends on not only the length and the cross section of the sample but the position inside the sample. These facts indicate that the figure of merit Z cannot be regarded as the peculiar material constant and should be precisely revised to be $Z=(\alpha^2/\rho\kappa)F(x, L, S)$, where x is a coordinates in the sample and S and L are the cross section and the length of the sample. The determination of the function $F(x, L, S)$ is still open but may be achieved by detail sample size and positional dependence measurements of ZT which are only possible by the Harman method. The present results are expected to be useful in the temperature control in the microscopic regions in which the thermoelectric properties in the neighborhood of the ends of the sample is inevitably important.

Figure captions

Fig. 1  Sample length L dependences of ZT at 300K in $(Bi,Sb)_2Te_3$ #1-1 and #2 samples.

Fig. 2 Temperature dependences of the dc and ac resistivity of the sample #1-2 for different voltage terminal length $L_V$.  In the inset $\Delta T/L_V$ versus $L_V$ plots are given to make the non-linear temperature distribution in the sample.

Fig. 3 Time dependences of the voltage by the ac current with 5mHz at room temperature for several $L_V$.  The amplitude of the thermal heat-flow is shown as a function of $L_V$ in the inset.

Fig. 4 $L_V/L$ dependences of ZT at 300K in the three thermoelectric materials.  Note that the horizontal axis for $Co_{0.9}(Pt,Pd)_{0.1}Sb_3$ and $\beta$-$Zn_4Sb_3$ materials is enlarged by twice as large as that for $(Bi,Sb)_2Te_3$.

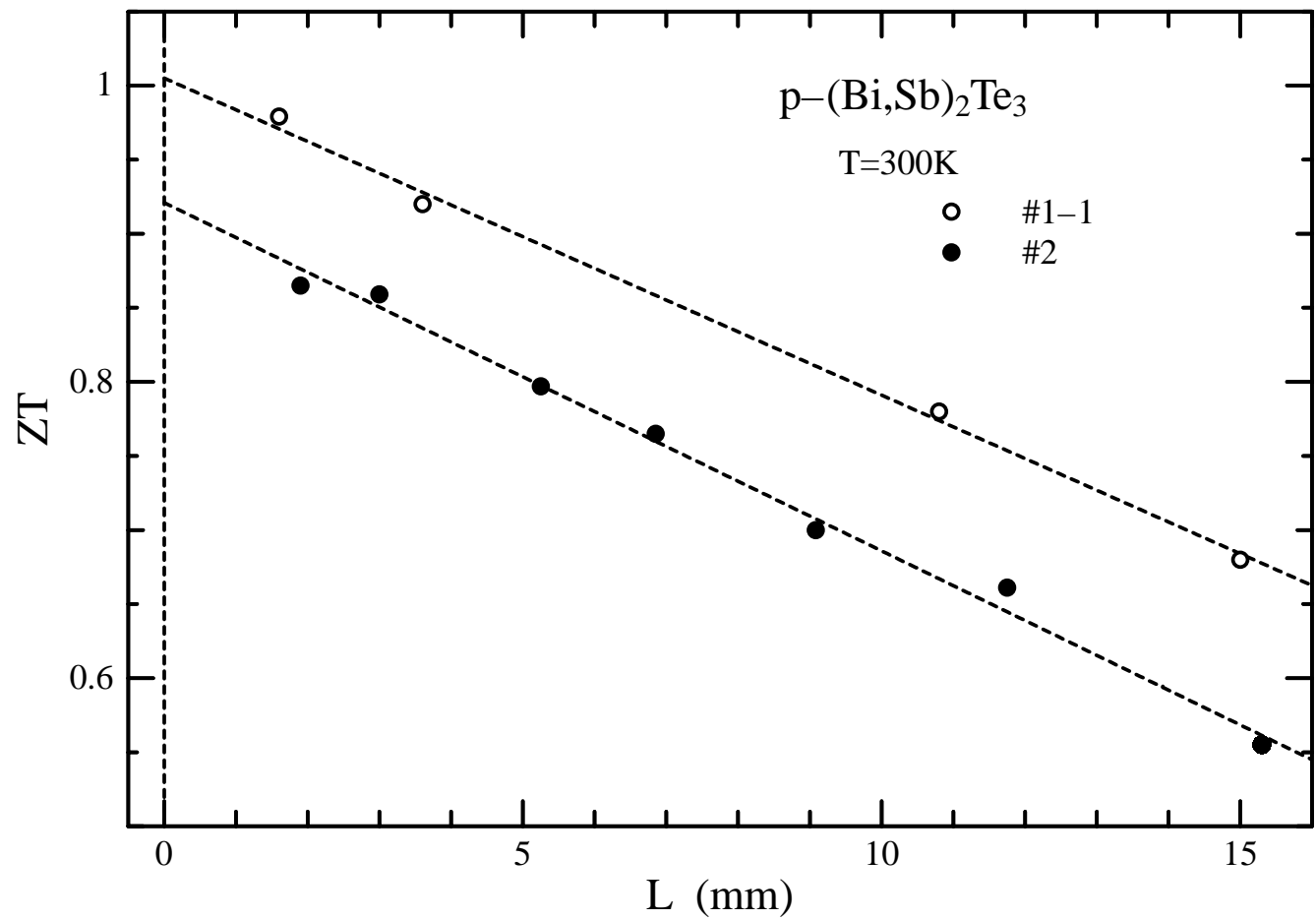

Fig. 1  H. Iwasaki et al.

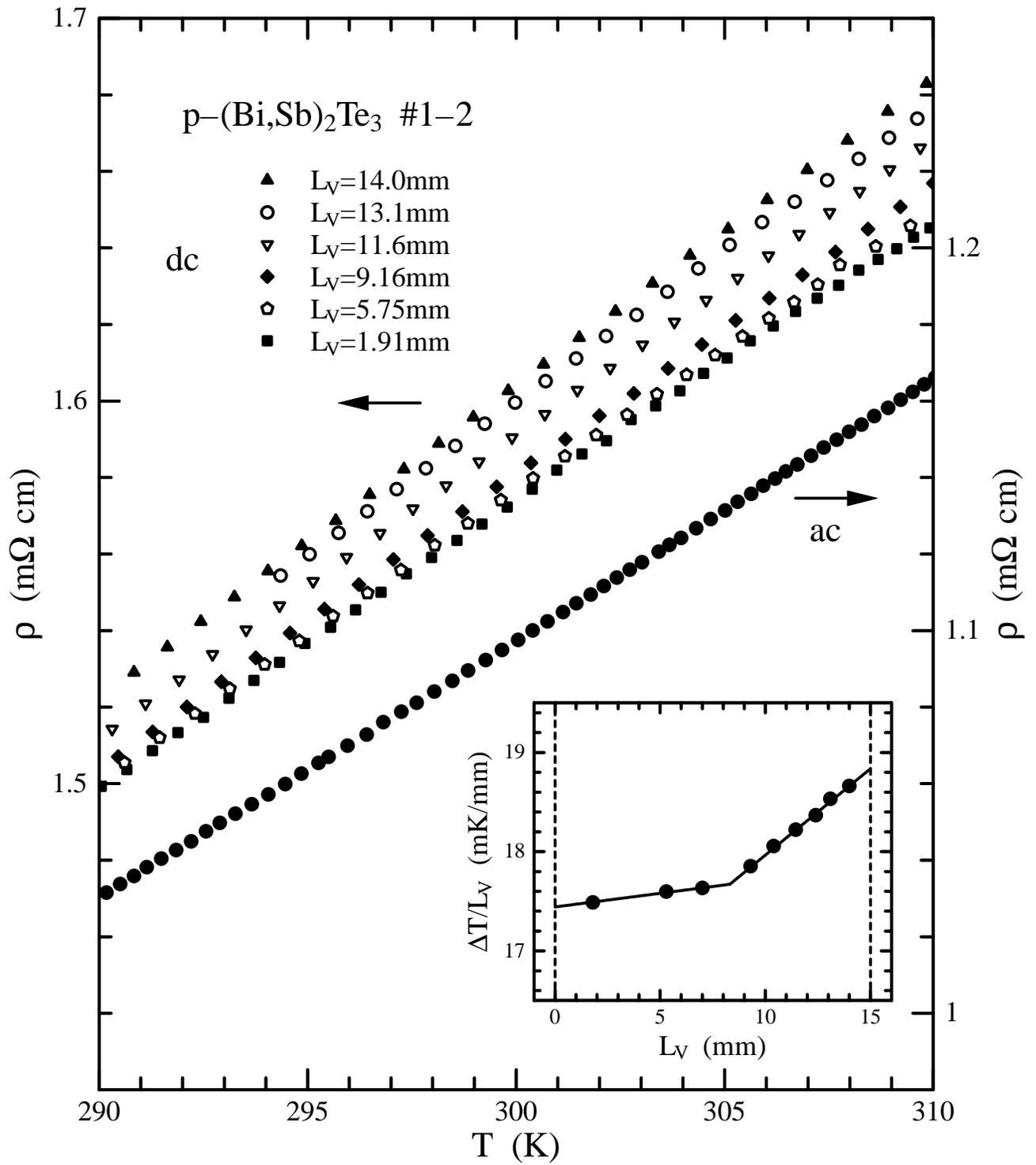

Fig. 2  H. Iwasaki et al.

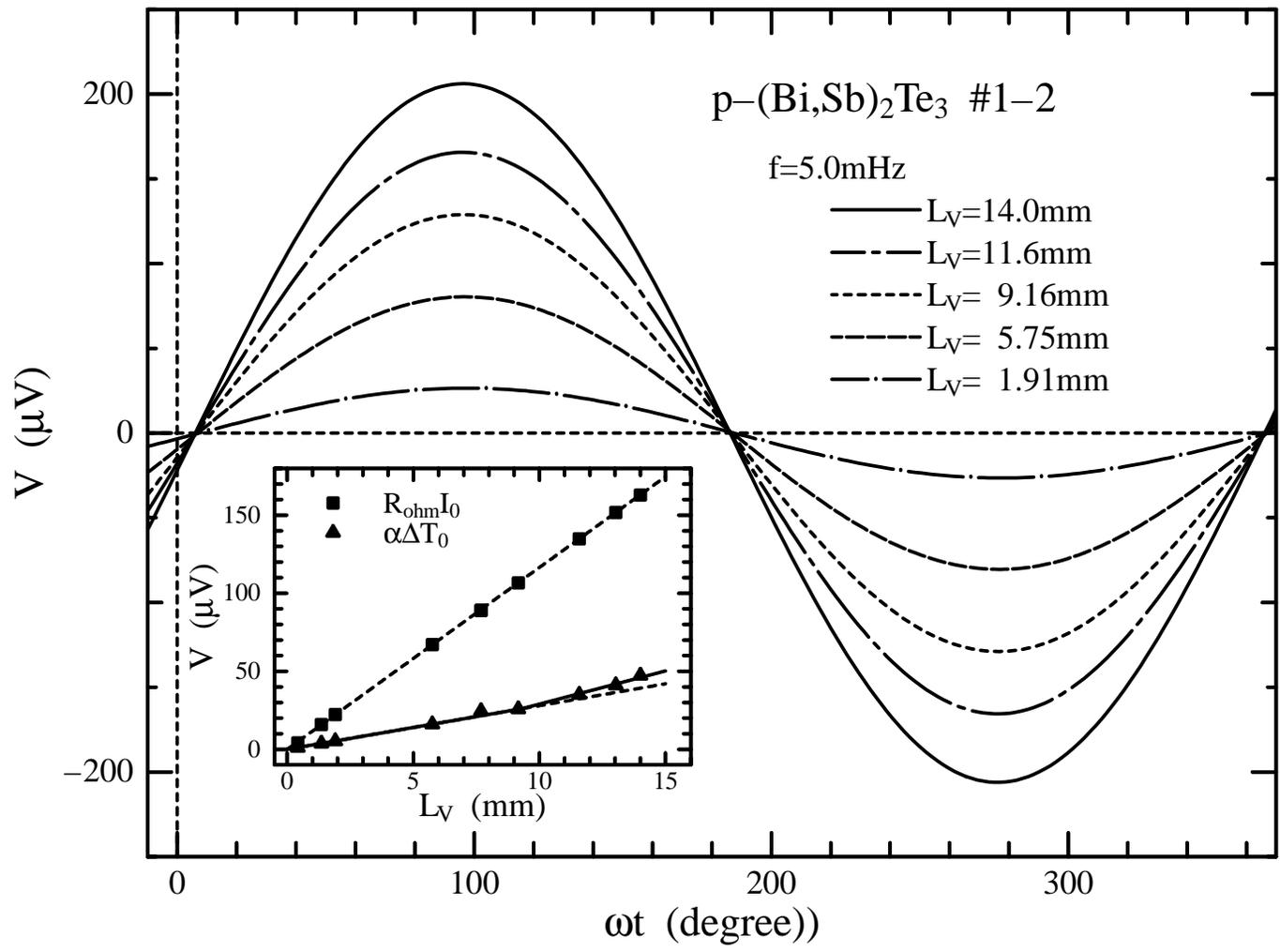

Fig. 3   H. Iwasaki et al.

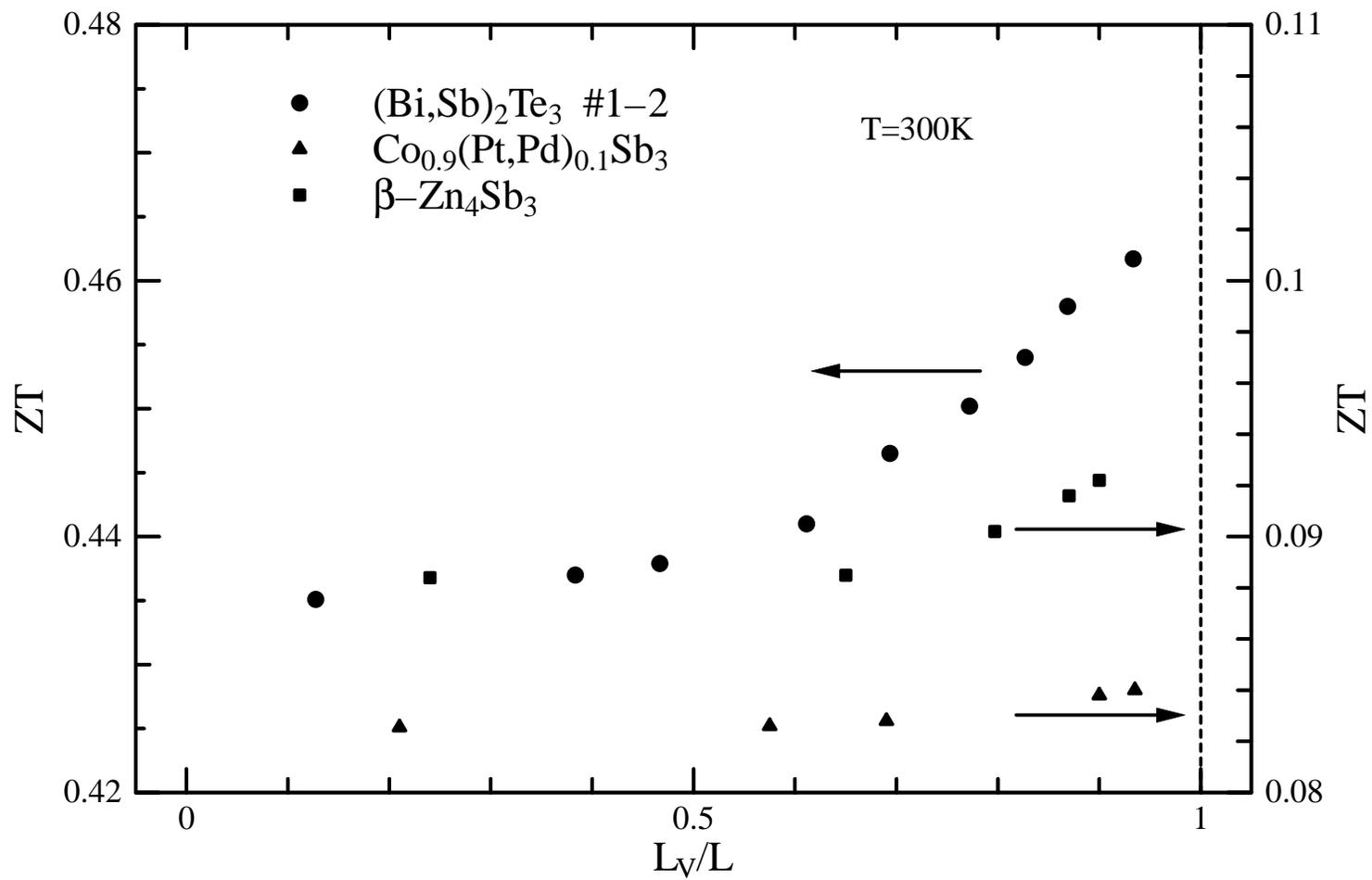

Fig. 4  H. Iwasaki et al.